\documentclass[runningheads]{llncs}
\usepackage[T1]{fontenc}
\usepackage{graphicx}
\usepackage[utf8]{inputenc} 
\usepackage[T1]{fontenc}    
\usepackage{hyperref}       
\usepackage{url}            
\usepackage{booktabs}       
\usepackage{amsfonts}       
\usepackage{nicefrac}       
\usepackage{microtype}      
\usepackage{xcolor}         
\usepackage{todonotes}
\usepackage{amsmath}
\usepackage{marvosym}

\usepackage{wrapfig}
\usepackage{subcaption}

\begin{document}
\title{DTU-Net: Learning Topological Similarity for Curvilinear Structure Segmentation}
%
%
\author{Manxi Lin\inst{1} \and
Kilian Zepf\inst{1} \and
Anders Nymark Christensen \inst{1} \and
Zahra Bashir \inst{2} \and
Morten Bo Søndergaard Svendsen \inst{3} \and Martin Tolsgaard \inst{3} \and
Aasa Feragen \inst{1}\textsuperscript{\Letter}
}
\authorrunning{Lin et al.}
%
\institute{$^1$~Technical University of Denmark, Kongens Lyngby, Denmark\\
\email{\{manli, afhar\}@dtu.dk}\\
$^2$~Slagelse Hospital, Copenhagen, Denmark\\
$^3$~Region Hovedstaden. Copenhagen, Denmark
}
\maketitle              
\begin{abstract}
Curvilinear structure segmentation is important in medical imaging, quantifying structures such as vessels, airways, neurons, or organ boundaries in 2D slices. Segmentation via pixel-wise classification often fails to capture the small and low-contrast curvilinear structures. Prior topological information is typically used to address this problem, often at an expensive computational cost, and sometimes requiring prior knowledge of the expected topology.

We present DTU-Net, a data-driven approach to topology-preserving curvilinear structure segmentation. DTU-Net consists of two sequential, lightweight U-Nets, dedicated to texture and topology, respectively. While the texture net makes a coarse prediction using image texture information, the topology net learns topological information from the coarse prediction by employing a triplet loss trained to recognize false and missed splits in the structure. We conduct experiments on a challenging multi-class ultrasound scan segmentation dataset as well as a well-known retinal imaging dataset. Results show that our model outperforms existing approaches in both pixel-wise segmentation accuracy and topological continuity, with no need for prior topological knowledge.      
\end{abstract}

\section{Introduction}
\begin{figure}[b]
  \centering
  \includegraphics[width=\linewidth]{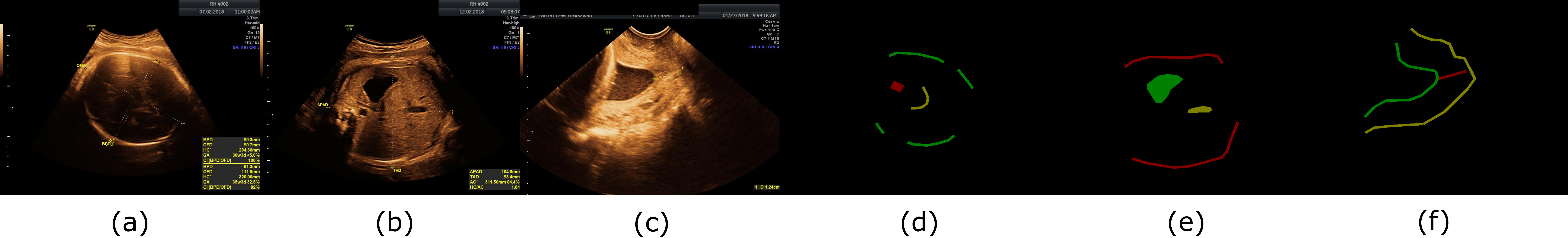} 
  \caption{Volumetric and curvilinear structures important for standard plane assessment in ultrasound of fetal head and abdomen, as well as maternal cervix are shown in (a)-(c), with manual segmentations in (d)-(f). Please zoom for details.
  }
  \label{fig:frontpage}
\end{figure}

Curvilinear structures represent thin and long objects in images, and their segmentation has extensive application in computer vision and medical imaging, e.g., segmentation of blood vessels~\cite{li2020iternet}, neurons~\cite{mou2021cs2}, or organ boundaries~\cite{hu2019topology}.

Segmenting curvilinear structures is difficult due to their long, thin shapes: Their connectivity can be completely altered by misclassifying a small number of pixels. This is particularly challenging in images with low contrast or resolution, where the structure blends more with the background. As a result, connectivity errors due to incorrectly splitting curves (we term these "\textbf{false splits}") and incorrectly connecting them (we term these "\textbf{missed splits}"), are very common.

In this paper, we consider curvilinear segmentation problems where missed splits are just as harmful as false splits. This is in contrast with parts of the existing literature~\cite{sasaki2017joint,clough2019topological}, which focuses on avoiding false splits or reproducing a prescribed topology by connecting components using prior knowledge. Our focus on missed splits is motivated by image quality assessment in fetal ultrasound~\cite{wu2017fuiqa}. Here, image quality depends on particular organs being completely visible in the image, making over-segmentation directly harmful. Examples of curvilinear structures important for fetal ultrasound quality assessment are shown in Figure~\ref{fig:frontpage}.

\begin{wrapfigure}{r}{0.28\textwidth}
\vspace{-8mm}
    \includegraphics[width=\linewidth]{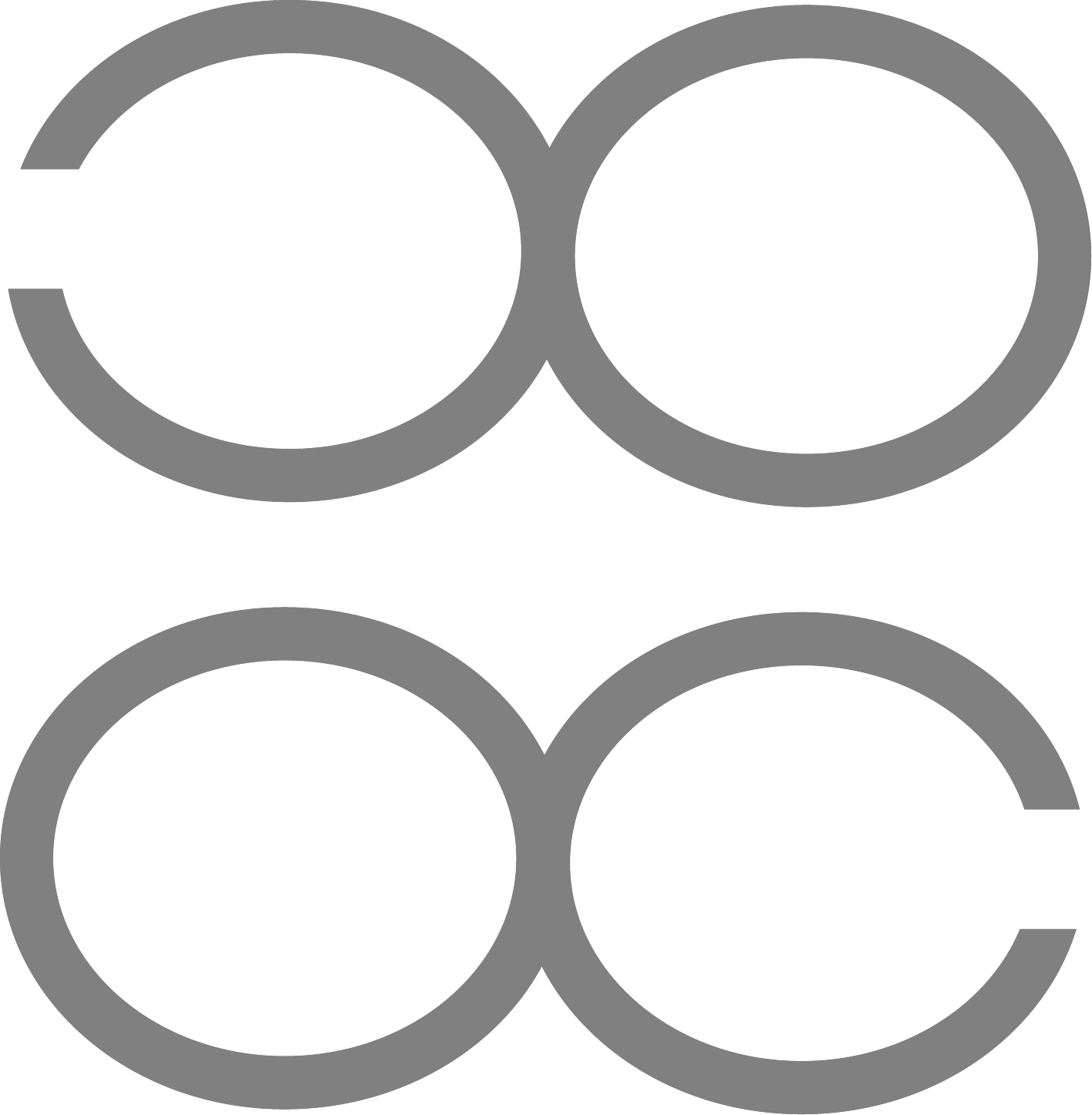}
  \caption{A globally topologically correct segmentation with local topological errors.}
  \vspace{-7mm}
  \label{fig:loctop}
\end{wrapfigure} \subsubsection{Curvilinear structure segmentation is a local topological problem.} Curvilinear structure segmentation is often described as a topological problem~\cite{shit2021cldice,clough2019topological}. Indeed, the topology of a subset of a topological space refers to those properties of the set that are invariant under continuous deformation. In curvilinear structure segmentation, the underlying task is to recover a curve that is visible as an elongated structure in the image. In such applications, it is often important to maintain the connectivity of the curve as accurately as possible. This is a topological problem: If $0 < a < b < 1$, then (the image of) a generic curve $c \colon [0, 1] \to \Omega$ is not topologically equivalent to (the image of) its restriction to an interval with a gap $(a, b)$ given by $c|([0, a] \cup [b, 1])$.

However, we want to emphasize that correct connectivity is a \emph{local} topological problem. To this end, note that topology is a global phenomenon, whereas image segmentation is inherently local. While topology, as quantified by Betti numbers, checks whether you have found the correct number of components or loops, this does not ensure a topologically correct segmentation.

To see this, consider the two curvilinear structures illustrated in Fig.~\ref{fig:loctop}, where the top illustrates a true segmentation and the bottom illustrates an estimated segmentation. These two segmentations are topologically equivalent, and their Betti numbers would be identical. They would also be rather similar in pixel-wise losses or measures. However, from the point of view of topology-preserving segmentation, the error made is considerable: The ground truth loop has not been detected, and the loop found in the segmentation is not supposed to be a loop. In other words, \textbf{the segmentation needs to be topologically correct in a local sense}. As we shall iterate throughout the paper, our modelling choices, as well as how we validate the topological correctness of segmented curvilinear structures, are directly linked to this modelling rationale.


\subsubsection{To obtain a data-driven curvilinear segmentation algorithm that is robust in the face of noisy data, we make the following contributions:}
\begin{enumerate}
    \item We build a dual-decoder neural network for curvilinear structure segmentation, generating predictions by softly fusing the two predictions produced by the two-stage prediction based on image texture and topology, respectively. The dual predictions give increased stability for low-quality images.
    \item Utilizing deep contrastive learning, we leverage the latent embedding similarity to learn topological information of the images in an end-to-end scheme, which does not rely on any prior information and produces self-supervised embeddings that respect the curvilinear structure topology. This data-driven approach to curvilinear structure segmentation is particularly scalable when faced with complex topological structures.
    
    \item We demonstrate our model on a challenging multi-class ultrasound segmentation task, and the open DRIVE retinal vessel segmentation dataset.
    
\end{enumerate}

\section{Related Work}

Image segmentation is usually formulated as pixel-wise classification based on image texture. However, this is not enough to capture complete curvilinear structures in the image, as small pixel-wise errors can lead to incorrect connectivity.  Early methods detected curvilinear structures by designing filters that target their geometrical properties, which are hard to detect at low contrast~\cite{frangi1998multiscale}. With the rise of deep learning, convolutional neural networks (CNNs), such as the U-net~\cite{ronneberger2015u}, have become the state-of-the-art for segmentation. 

For curvilinear structure segmentation, one branch of research uses persistent topology to~\cite{clough2019topological,hu2019topology,hu2021topology} to train networks to explicitly avoid breaking topological structure. A caveat with such approaches, especially for complex topologies, is that they require expensive topological computation. Following a similar, but less expensive route, Shit et al.~\cite{shit2021cldice} define a clDice loss that  encourages the segmentations to maintain correct topology based on their soft skeletons.

Another branch of work~\cite{mou2021cs2,li2020iternet} designs delicate backbones and schemes to capture the details of curvilinear features. CS-Net~\cite{mou2021cs2} introduces a self-attention mechanism that deals with multiple curvilinear types in a unified manner. IterNet~\cite{li2020iternet} concatenates U-Net and multiple mini U-Nets to incorporate the structural redundancy. Nevertheless, these methods, which simply learn semantic labels per pixel, often fail to preserve the connectivity of objects. TR-GAN~\cite{chen2020tr} is a generative adversarial network that improves artery/vein classification of vessels by improving segmented vessel connectivity. A topology ranking discriminator along with a contrastive loss encourages the network to generate segmentation with the expected topology ranking. The topological features learned by a pre-trained network are only involved in the computation of cost functions to preserve the image topology. These methods have in common that they are not designed to handle low-quality images. The pre-trained features used in~\cite{chen2020tr,mosinska2018beyond} are less well suited for such problems, and we find experimentally that architectures such as~\cite{li2020iternet} struggle with low-quality images. In this paper, we, instead, explicitly model the types of errors for which we wish the network to be robust.

A different alternative is given by post hoc approaches, which correct the broken predictions from an independent segmentation model. Sasaki et al.~\cite{sasaki2017joint} propose an auto-encoder that detects and paints the gaps in images automatically. These models suffer from the clear disadvantage that they can only repair false splits -- they are unable to handle missed splits.

All the previously mentioned models have in common that they aim to preserve topology and encode image textures in a single encoding stage. We find that in particular for images with low contrast or resolution, combining these two tasks at the same stage may make the model hard to tune. Thus, in this paper, we split the topology-preserving curvilinear segmentation into two simpler sub-tasks: First learning texture information, and next learning topological information from the texture-based coarse segmentation.

\section{Method}

\subsection{Decoupling Topology-preserving Curvilinear Segmentation} 
\label{sec: revisit}

\begin{figure}[b]
  \centering
  \includegraphics[width=\linewidth]{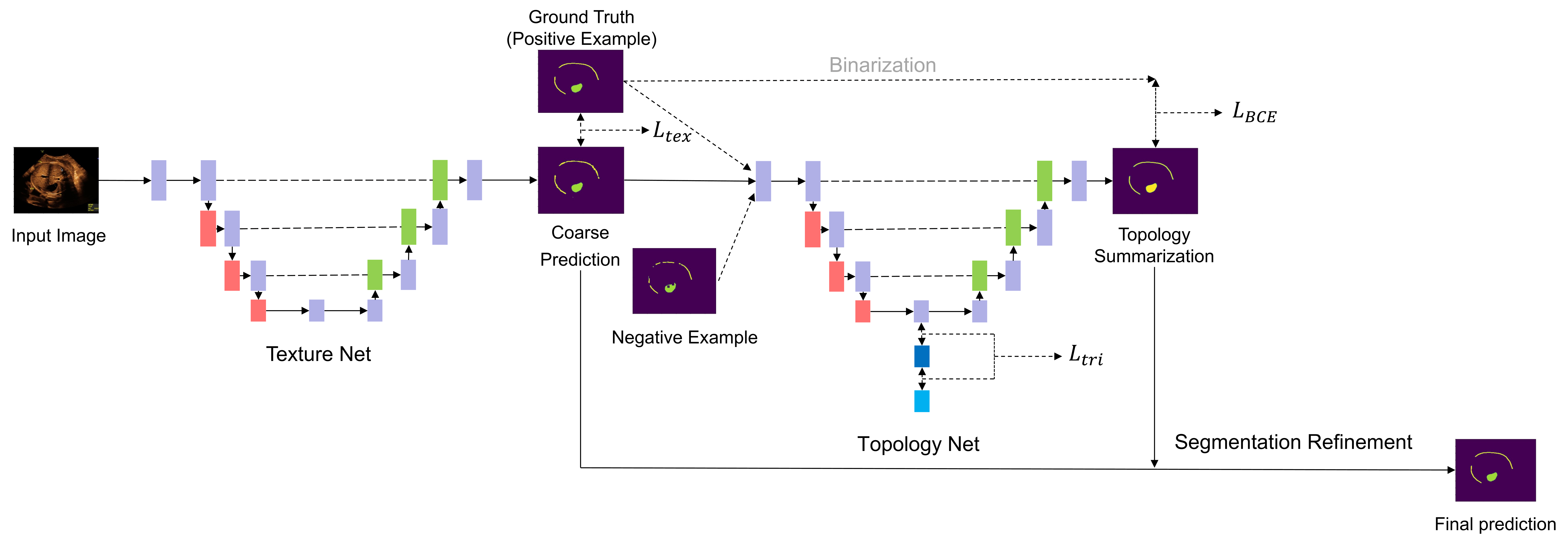} 
  \caption{Network architecture of DTU-Net. The dotted arrows refer to operations that are only performed at the training stage. Please zoom in for details. }
  \label{fig:net}
  \vspace*{-\baselineskip}
\end{figure}

We first summarize our proposed \textbf{D}ual-decoder and \textbf{T}opology-aware \textbf{U-Net} (\textbf{DTU-Net}) architecture shown in Figure~\ref{fig:net}. Our DTU-Net includes two lightweight mini U-Nets, with fewer layers than the original U-Net, referred to as the \emph{texture net} and the \emph{topology net}. The mini U-Net architecture is the same as that of the mini U-Nets from IterNet~\cite{li2020iternet}. The texture net follows the learning scheme of most general segmentation tasks - encoding image texture information and giving pixel-wise predictions. The coarse prediction from the texture net is further fed to the topology net for topology preservation. The topology net is another mini U-Net, where the encoder learns topological features using a self-supervised triplet loss, and the decoder summarizes the topology in a binary segmentation map indicating image foreground and -background. The texture-based and topology-based predictions are further softly fused to obtain the final segmentation.

Note that in practice, the texture net can be replaced with any pre-trained or randomly-initialized segmentation network. This means that our model can also be employed as an easy-to-use plugin to extend to powerful general segmentation models. Our use of lightweight mini-U-nets is motivated by the ultrasound image quality assessment task, where we aim for segmentations to serve as real-time feedback on image quality during scanning, with limited computational resources.

\subsection{Topology Net}
\label{sec: topo}
\subsubsection{Learning topological embeddings with a triplet loss.} Recent approaches to curvilinear structure segmentation~\cite{chen2020tr,mosinska2018beyond} adopt a pre-trained VGG19 network to extract high-level image topological features, where contrastive losses penalize segmentations with different topological features from the ground truth. In these methods, topological information is extracted by models pre-trained on natural images, which is agnostic to the task. In addition, during training, only the feature differences are considered instead of the features themselves. We, instead, integrate contrastive learning into the latent embedding of a standard segmentation network using self-supervised learning to learn topological consistency.

Contrastive learning is often used to extract effective instance features for downstream tasks by producing embeddings that keep similar objects close, and push dissimilar objects apart. Here, we adopt a triplet loss in the bottleneck of the topology net, enabling self-supervised learning of image topology features.

The triplet loss compares an anchor point to a positive and negative example~\cite{schroff2015facenet}. The loss function penalizes the difference between the anchor's distance to the positive and negative examples, respectively. In our formulation, the image is passed first through the texture net $\Theta$ to obtain a coarse segmentation, and further through the encoder $\Psi$ of the topology net, where the result $\Psi(\Theta(I))$ is used as an \emph{anchor}. The positive example $\Psi(G)$ is obtained by passing the ground truth segmentation $G$ through the topology net encoder, and the negative example $\Psi(\hat{G})$  is similarly obtained from a segmentation $\hat{G}$, which has been corrupted to alter its topology. As detailed in the next section, we generate positive and negative examples from segmentation ground truth online during training and do not need additional annotations.

Sequentially, the triplet loss is defined as:
\begin{equation}
    L_{tri}(\Psi(\Theta(I)), \Psi(G), \Psi(\hat{G})) = max(d^{+} - d^{-} + \tau, 0)
    \label{triplet}
\end{equation}
where $d^{+}$ is the averaged Euclidean distance of $\Psi(\Theta(I))$ and $\Psi(G)$, $d^{-}$ is the distance of $\Psi(\Theta(I))$ and $\Psi(\hat{G})$ and $\tau$ is an empirically-set margin. By minimizing the triplet loss during training, the network forces the encoded features closer to those of the ground truth segmentation mask whose topology is correct, while pushing it further away from the embedding of the corrupted segmentation masks.  

\subsubsection{Self-supervised training of the triplet loss.} We generate corrupted masks from the ground truth mask by randomly breaking the image topology. Synthetic false splits in the corrupted masks are generated as follows: We assume that false splits can happen anywhere on the curves when the CNNs fail to recognize the objects. Therefore, segmentation labels are split into small patches, randomly removing $\lambda$ of the patches containing foreground information, where $\lambda \in [0,1]$ is a hyper parameter controlling the degree of corruption. 

Next, we also generate corrupted masks with synthetic missed splits: We deem the missed splits tend to appear in regions with similar texture features, put simply in raw images, the pixel values. We compute the average $m$ and standard deviation $d$ of the foreground pixel values in the image (foreground pixels are given in the ground truth). To generate synthetic missed splits, a fixed percentage $\lambda$ of the pixels with value within $[m-d, m+d]$ are selected randomly and assigned the label of the most similar foreground pixel. 

In our implementation, adding missed splits and false splits is done sequentially. The introduction of missed splits brings not only incorrect connections between curves but also noise, which harms the topology as well.

During inference, the encoder only forwards $\Psi(\Theta(I))$. The triplet loss in Eq.~\ref{triplet} quantifies the pixel-wise feature difference. We also merge the global topological feature into the embedding by a SENet~\cite{hu2018squeeze}, which takes the average of the learned feature map and encodes them into attentive scores via a multi-layer perceptron. The scores are then multiplied by the original feature map. 

\subsubsection{Fine-to-coarse topology summarization. }
 

In the DTU-Net, the texture net learns low-level information directly from raw images, which is effective in semantic segmentation tasks. Our topology net, however, abstracts the information into high-level topological features, which are less applicable for texture recognition tasks due to the loss of texture details. We expect the topology net to summarize the information from object connectivity, and contribute to the final segmentation. Consequently, we ask the topology net a binarized question: Based on the learned topology, should the structure be connected or not? For multi-class segmentation tasks such as in our ULTRASOUND dataset, these binary predictions are sequentially employed to refine the multi-class semantic segmentation from the texture net. 
 
 

\subsection{Segmentation Refinement and Joint Feature Learning}
\label{sec:fuse}
Our DTU-Net consists of a texture network and a topology network. We denote the texture network (a mini U-Net), the encoder, and the decoder of the topology network (another mini U-Net) as $\Theta(\cdot)$, $\Psi(\cdot)$, and $\Omega(\cdot)$ respectively. Each part of the network can be trained separately, as with the post hoc methods. We train the network in a joint way, where the gradient backpropagation from the topology net can also update the parameters in the texture net. Denoting input images $I$ and segmentation ground truth masks $G$, DTU-Net is trained by a unified loss 
\begin{equation}
    L_{DTU}(I, G) = L_{tex}(\Theta(I), G) + L_{BCE}(\Omega(\Psi(\Theta(I))), \bar{G}) + L_{tri}
\label{eq: loss}
\end{equation}
where $L_{tex}$ is the pixel-wise segmentation loss for the texture net, $L_{BCE}$ is the binary cross entropy loss tuning the prediction from the topology net, $L_{tri}$ is the triplet loss in Eq.~\ref{triplet}, $\bar{G}$ and $\hat{G}$ refers to the ground truth binarized and corrupted by us. The loss function also indicates that our DTU-Net does not need additional annotations and learns image topology in a self-supervised way. 

As the topology net and the texture net view the segmentation problem from different points of view, we fuse their predictions into a final prediction -- this avoids sensitivity to the topology net sometimes being overconfident. Inspired by ensemble learning~\cite{schapire1999brief}, we fuse the two predictions using weighted summation, where the hyper parameter $\omega$ indicates our confidence in the texture-based prediction. For each pixel, given $p_{top}$ the pixel-wise probability of the topological prediction, and $p_{tex}^{(i)}$ the pixel-wise probability of the $i$-th category from the textural prediction, the two predictions are simply fused by:         
\begin{equation}
    p^{(i)}_{final} = 
    \begin{cases}
    (1-\omega) (1-p_{top}) + \omega p^{(0)}_{tex}, i=0;\\ 
    \frac{(1-\omega)p_{top} + \omega \sum_{i=1}^{c}p^{(i)}_{tex}}{\sum_{i=1}^{c}p^{(i)}_{tex}}p^{(i)}_{tex}, i=1,2,...,c
    \end{cases}
    \label{eq:weight}
\end{equation}
where $c$ is the total number of categories in the semantic segmentation task, $p^{(i)}_{final}$ is the pixel-wise final prediction, the $0$-th category means the background.

\section{Experiments}

\subsubsection{Baseline models.} We compare our model against a mini U-net trained with the clDice loss function, as well as with the persistent homology-based TopoLoss~\cite{hu2019topology}; the data-driven IterNet~\cite{li2020iternet} and the inpainting network of~\cite{sasaki2017joint}. The input of the inpainting network is the prediction from a mini U-Net trained by a focal loss.  

The clDice loss function preserves topological information by learning object skeletons, which, however, does not guarantee accurate segmentation. In the original paper~\cite{shit2021cldice}, this is handled by mixing the clDice metric with another pixel-wise classification loss, more precisely the soft dice loss with hyper-parameter $\beta$:
\begin{equation}
    L_{clDice} = \beta (1 - clDice) + (1-\beta) (1-softDice)
\end{equation}

In our experiments, and in particular on the more challenging ULTRASOUND dataset, we found the clDice loss function challenging to tune. The experimental setup from ~\cite{shit2021cldice} did not converge with of the suggested values of $\beta = 0$ or $0.5$ (see Table\ref{tab:us_cldice}). To improve convergence, we replaced softDice with focal loss~\cite{lin2017focal}:
\begin{equation}
    L_{clDice} = \alpha (1 - clDice) + (1-\alpha)L_{focal}.
\end{equation}
Even here, however, the performance was unstable with the choice of $\alpha$, as can be seen from Table~\ref{tab:us_cldice}, and we choose to compare to the optimal observed $\alpha=0.2$ in Table~\ref{tab:us}. For the DRIVE dataset, we choose $\alpha=0.5$ following~\cite{shit2021cldice}. 

\begin{table}[t]
  \caption{Results on the ULTRASOUND dataset. Note that cIoU and cmIoU refer to IoU and mean class IoU, respectively, for curvilinear structures, whereas vIoU and vmIoU refers to volumetric structures.} 
  \label{tab:us}
  \centering
  \begin{tabular}{ll|l|l|llll}
    \toprule
    Model & Loss & Frechet & Betti error & vIoU & vmIoU & cIoU & cmIoU
    \\    
    \midrule
    mini U-Net     & clDice & 0.7801  & 0.2635  & 97.95 & 29.84 & 31.41 & 20.41\\
    mini U-Net     & TopoLoss & 0.9653 & 0.2606 & 98.07 & 26.29 & 25.89 & 18.56
    \\
    IterNet     & focal      & \textbf{0.4329} & 1.0963  & 98.20  & 17.67 & 5.76 & 3.70  
    \\
    inpainting     & MSE     & 0.8189 & 1.1179  & 93.88 & 23.93 & 40.15 & 24.37  
    \\
    DTU-Net     & Eq.~\ref{eq: loss} & 0.5818 & \textbf{0.0974}  & \bf 99.51 & \bf 44.33  & \bf 72.30 & \bf 52.92
    \\
    \bottomrule
  \end{tabular}
\end{table}

\begin{table}[t]
  \caption{Segmentation results on the DRIVE dataset.}
  \label{tab:drive}
  \centering
  \begin{tabular}{l|l|l|l|l}
    \toprule
    Model & Loss & Frechet & Betti error & IoU \\
    \midrule
    mini U-Net     &  clDice    & 3.6850 & 1.0890 & 67.95  \\
    mini U-Net     &  TopoLoss    & 3.6070  & 0.9232 &  70.52 \\
    IterNet     & focal        & 3.5335 & 1.0039 & 70.35  \\
    inpainting    & MSE      &  3.9001 & 1.3824 &  65.54 \\
    DTU-Net     & Eq.~\ref{eq: loss}  & \bf 2.9316 & \bf 0.8597 & \bf 73.86 \\
    \bottomrule
  \end{tabular}
\end{table}

\subsubsection{Tasks.} We validate our algorithm on two different segmentation tasks. 

\begin{table}[]
    \caption{(a) clDice model selection on ULTRASOUND. The first two rows show the models suggested in the original paper~\cite{shit2021cldice}; these did not converge. The next rows show results for various parameters using the focal loss in place of the soft dice. (b) Training times on ULTRASOUND. TopoLoss and clDice were used with a mini U-Net; the inpainting included two stages.}
    \begin{subtable}{.65\linewidth}
      \caption{}
      \label{tab:us_cldice}
      \centering
            \begin{tabular}{ll|l|l|llll}
    \toprule
    \cmidrule(r){1-2}
    $\alpha$ & $\beta$ & Frechet & Betti error  & vIoU & vmIoU & cIoU & cmIoU\\  
    \midrule
        0  &  0.5 & - & 9.8019  & 2.14 & 0.33 & 1.53  & 1.21 \\
    0 & 1  & - & 13.93  & 3.72 & 0.62 & 2.91 & 1.80 \\
        \midrule
    0  &  0  & 0.8418 & 0.4572  & 97.70 & 29.42 & \bf 32.99 & \bf 23.14 \\
    0.1  & 0  & 0.9500  & 0.3800   & 97.74 & 28.13 & 26.34 & 17.72\\ 
    0.2 & 0 & \bf 0.7801 & \bf 0.2635   & \bf 97.95 & \bf 29.84 & 31.41 & 20.41\\
    0.3  & 0  & 1.0798  & 0.4128  & 97.70 &  28.75 & 29.79 & 21.07\\ 
    0.4  & 0  & 1.0727  & 0.3873  & 97.44 & 29.30 & 25.56 & 18.09\\ 
    0.5  & 0  & 0.8969  & 0.3264  & 97.06 & 26.69 & 22.76 & 15.00\\ 
    \bottomrule
  \end{tabular}
    \end{subtable} \hfil
    \begin{subtable}{.3\linewidth}
      \centering
        \caption{}
        \label{tab:runtime}
  \begin{tabular}{l|l}
    \toprule
Model &	Time/hours\\
\midrule
TopoLoss & 	43.74\\
IterNet & 20.02\\
clDice &	6.91\\
inpainting 	&7.47\\
DTU-Net & 17.2\\
\bottomrule
\end{tabular}
    \end{subtable} 
\end{table}

The ULTRASOUND dataset poses a challenging fetal ultrasound multiclass segmentation task including both volumetric and curvilinear structures. These images, collected from the Danish national fetal ultrasound screening database, are obtained a typical 3rd-trimester growth scans, where four standard ultrasound planes (head, abdomen, femur, cervix) are acquired to assess preterm birth and fetal weight. Whether the images are "standard" depends on the visibility and pose of certain anatomical organs, several of which appear as curvilinear structures in the image. It is important to accurately detect whether the entire structure is present, as well as measuring its dimensions. Our motivating downstream application requires segmenting multiple different organs in real time while scanning; we have trained our models to segment both the volumetric and curvilinear structures. There are 2088 images of resolution $960\times 720$, divided into $80\%$ for training, $10\%$ for validation, and $10\%$ for testing. The dataset includes 6 curvilinear (cervix canal outline, cervix outer boundary, cervix inner boundary, outer skin boundary, outer bone boundary, thalamus) and 7 volumetric segmentation categories (bladder. femur bone, stomach bubble, umbilical vein, kidney, fossa posterior, cavum septi pellucidi). There are 468 femur, 531 abdomen, 686 head, and 403 cervix images. Images are reshaped to $224\times 288$. 

The DRIVE retinal segmentation dataset~\cite{drive} induces a binary segmentation task. We used $16$ images for training and $4$ for testing. During training, the images are randomly cropped into $256 \times 256$ patches, while in testing patches are obtained with a $128 \times 128$ window sliding with a stride of $64$. 

\subsubsection{Experimental settings.} The texture net is trained with a combined dice and focal loss as $L_{tex}$ in Eq.~\ref{eq: loss}. Triplet loss examples are generated incrementally by reducing $\lambda$ from 50\% to 10\% during training, with $\omega= 0.5$ and $\tau = 0.1$. 

\subsubsection{Evaluation metrics. } To measure geometric affinity between curves, we employ the Frechet distance~\cite{alt1995computing}, which measures the extent to which the two curves remain close to each other throughout their course. We find this more suitable than e.g.~accuracy, which is sensitive to small shifts or differences in curve thickness.

To measure the local topological correctness of the segmented structures, we apply a sliding window, in which we compare the number of components in the prediction and the ground truth annotation, respectively. This corresponds to a local Betti number computation. For each image, we compute a topological error (Betti error) which is the average difference in Betti numbers between ground truth and prediction over all patches seen, similar to what is done in~\cite{hu2019topology}. 

For completeness, we also include the voxel-wise intersection over union (IoU), and mean class IoU for the multi-class segmentation task. For the ULTRASOUND dataset we report these both for the curvilinear and volumetric structure classes.

\begin{table}[t]
  \caption{Ablation study on different components of DTU-Net.}
  \label{tab:ablation}
  \centering
  \centering
    \begin{tabular}{l|l|l|llll}
    \toprule

    Model & Frechet & Betti error & vIoU & vmIoU & cIoU & cmIoU \\
     \toprule
    w/o triplet loss & 0.6605 & 0.1056 & 98.95  & 31.56 & 66.45  & 41.21 \\
    $\omega=0$ &  0.5863 & 0.1023 & 98.90  & 43.11 & 70.43 & 52.33 \\
    $\omega=1$   & 0.8601 & 0.1683 & 98.12  & 14.15 & 30.88 & 22.20\\ 
    \midrule
    DTU-Net  & \bf 0.5818 & \bf 0.0974  & \bf 99.51 & \bf 44.33 & \bf 72.30 & \bf 52.92 \\
    \bottomrule  
  \end{tabular}
\end{table}

\begin{table}[t]
  \caption{Ablation on Model size}
  \label{tab:ablation_size}
  \centering
  \begin{tabular}{ll|l|l|llll}
    \toprule
    Model & Size & Frechet & Betti error & vIoU & vmIoU & cIoU & cmIoU
    \\    
    \midrule
    full U-Net & 131.8M & 0.5849 & 0.1383 & 98.53 & 31.09 & 32.58 & 23.22  
    \\
     \midrule
    DTU-Net & 65.5M & \textbf{0.5818} & \textbf{0.0974} & \textbf{99.51} & \textbf{44.33} & \bf 72.30 & \bf 52.92 
    \\
    \bottomrule
  \end{tabular}
\end{table}

\subsubsection{Results.} 
\begin{figure}[t]
  \centering
  \includegraphics[width=\linewidth]{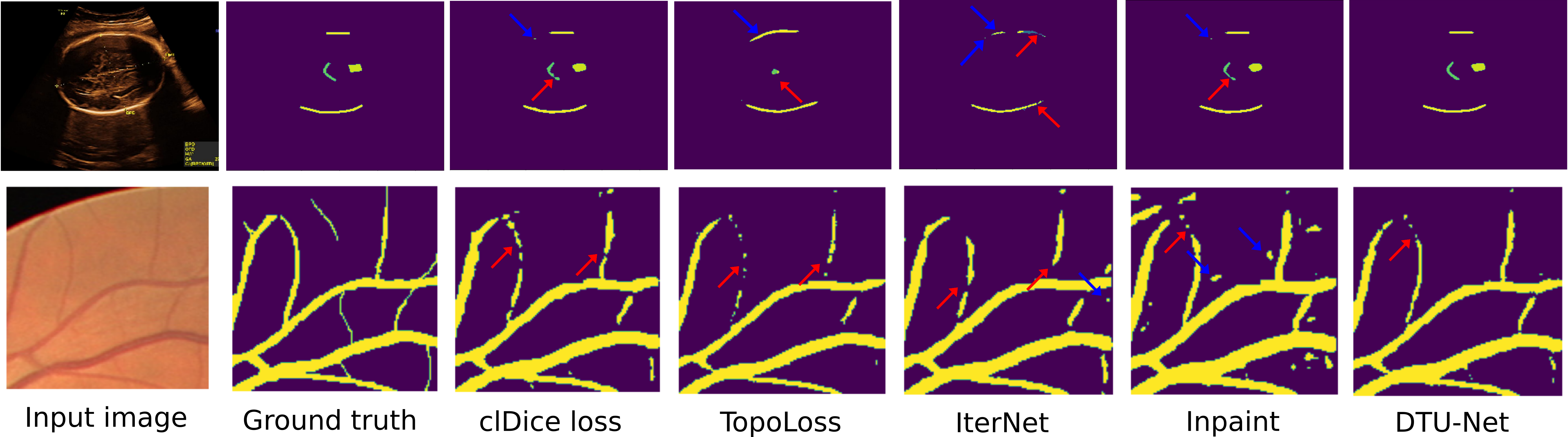}
  \caption{Segmentation of examples from ULTRASOUND and DRIVE. Blue arrows point at missed splits while red arrows point at false splits in the image.}
  \label{fig:seg_results}
\end{figure}

Results for the ULTRASOUND dataset are found in Table~\ref{tab:us}, and for the DRIVE dataset in Table~\ref{tab:drive}, with example segmentations in Figure~\ref{fig:seg_results}. The performance of the model used as the input for the post hoc inpainting on the ULTRASOUND dataset can be found in Table~\ref{tab:us_cldice} with $\alpha$ and $\beta = 0$. 

\subsubsection{Running time.} Table~\ref{tab:runtime} compares training times on ULTRASOUND. The models were trained on a Quadro RTX 6000 GPU with identical training settings. Note how DTU-Net has a considerably lower training time than TopoLoss, which is the closest competitor on the DRIVE dataset. While clDice and inpainting have the lowest runtimes, these do not compete in performance (Tables~\ref{tab:us} and~\ref{tab:drive}).

\subsubsection{Ablation study.} We conducted an ablation study over different components of the DTU-Net on the ULTRASOUND dataset, see Table~\ref{tab:ablation}. We see that removing the triplet loss from Eq~\ref{eq: loss} leads to increased Betti error. We also evaluate the effectiveness of feature fusion. Specifically, $\omega$ in Eq~\ref{eq:weight} is set to be 0 and 1 respectively, meaning that the prediction of foreground and background is dominated by the topology net or the texture net. To prove that our model performance is not merely due to the increase of learnable parameters, we also include a comparison with a larger U-Net trained with the clDice loss ($\alpha=0.2$), which has roughly two times the model size of ours (Table~\ref{tab:ablation_size}).

\section{Discussion and Conclusion}
We have designed a framework for curvilinear structure segmentation which is trained, in a data driven way, to avoid both false and missed splits. Our method is built around a dual-decoder network, which first makes a rough segmentation based on image texture, and next refines this optimizing for predicting correct topological structures. Here "correctness" also emphasises whether the structure is visible in the image. As the second network takes the predicted foreground probabilities as input, it learns to enhance those responses that form curvilinear structures. This is obtained using self-supervised contrastive learning, to teach the network the difference between a weak curvilinear response, and no response. On the highly challenging ULTRASOUND dataset, we see that our method outperforms all other methods. This is natural as most methods are developed using easier segmentation tasks such as e.g.~vessel segmentation, and they are not optimized to handle the challenging nature of ultrasound data. 

However, our method also outperforms baselines on the well-known DRIVE dataset. Here, we hypothesize that the data driven approach may have an easier time learning to handle complex topologies. It also has an advantage over persistent homology based methods, as it explicitly seeks to preserve local topology, whereas persistent homology based methods achieve locality by preserving Betti numbers within patches. This means that the locality scale becomes a parameter, which is additionally constrained by computational demands. This gives the DTU-net an advantage both in terms of choice of scale (it is inherently robust to scale) and in terms of computational time, as also shown by our experiments.

One important difference between methods is the number of parameters used. We have designed most methods around the fixed size mini U-net. Since our own method has the additional topology network, it de facto has more parameters than the mini U-net trained with clDice. To asses whether this difference in parameters was unfair, we also compared our method to a deeper U-net (with more parameters than our full architecture), trained with clDice. However, the results (see Table~\ref{tab:ablation_size}) show that we still outperform clDice on the ULTRASOUND dataset even when it is given the advantage of more parameters. The IterNet architecture also has more parameters than ours, but is not competitive.

\textbf{Limitations.} Our method is naturally limited by our downstream need for real-time inference, which is why we rely on mini U-Nets rather than more complex models. Allowing larger models might change the outcomes for competing methods, although this was not the case for clDice on the ULTRASOUND dataset. At the same time, our architecture's two mini U-Nets makes our method more expensive, also at inference time, than single mini U-Nets trained with topological loss functions. Finally, while the data-driven contrastive learning is clearly helpful in learning topological structure, it does not come with theoretical guarantees.

\textbf{Summary.} We have proposed a data-driven framework for segmenting curvilinear structures, and shown that it outperforms the baselines on DRIVE, and by far on the challenging ULTRASOUND dataset. We find it interesting that all baseline models struggle on the ultrasound data, and see this as a cue for the community to include more challenging segmentation tasks when developing methods for curvilinear structure segmentation.

\bibliographystyle{splncs04}
\bibliography{ref} 

\begin{thebibliography}{10}
\providecommand{\url}[1]{\texttt{#1}}
\providecommand{\urlprefix}{URL }
\providecommand{\doi}[1]{https://doi.org/#1}

\bibitem{alt1995computing}
Alt, H., Godau, M.: Computing the {F}r{\'e}chet distance between two polygonal
  curves. International J.~Computational Geometry \& Applications
  \textbf{5}(01n02),  75--91 (1995)

\bibitem{chen2020tr}
Chen, W., Yu, S., Wu, J., Ma, K., Bian, C., Chu, C., Shen, L., Zheng, Y.:
  {TR-GAN}: topology ranking {GAN} with triplet loss for retinal artery/vein
  classification. In: International Conference on Medical Image Computing and
  Computer-Assisted Intervention. pp. 616--625. Springer (2020)

\bibitem{clough2019topological}
Clough, J., Byrne, N., Oksuz, I., Zimmer, V.A., Schnabel, J.A., King, A.: A
  topological loss function for deep-learning based image segmentation using
  persistent homology. IEEE Transactions on Pattern Analysis and Machine
  Intelligence  (2020)

\bibitem{frangi1998multiscale}
Frangi, A.F., Niessen, W.J., Vincken, K.L., Viergever, M.A.: Multiscale vessel
  enhancement filtering. In: International conference on medical image
  computing and computer-assisted intervention. pp. 130--137. Springer (1998)

\bibitem{hu2018squeeze}
Hu, J., Shen, L., Sun, G.: Squeeze-and-excitation networks. In: IEEE CVPR
  (2018)

\bibitem{hu2021topology}
Hu, X., Wang, Y., Li, F., Samaras, D., Chen, C.: Topology-aware segmentation
  using discrete morse theory. In: ICLR (2021)

\bibitem{hu2019topology}
Hu, X., Li, F., Samaras, D., Chen, C.: Topology-preserving deep image
  segmentation. Advances in Neural Information Processing Systems  \textbf{32}
  (2019)

\bibitem{li2020iternet}
Li, L., Verma, M., Nakashima, Y., Nagahara, H., Kawasaki, R.: Iternet: Retinal
  image segmentation utilizing structural redundancy in vessel networks. In:
  Proceedings of the IEEE/CVF Winter Conference on Applications of Computer
  Vision. pp. 3656--3665 (2020)

\bibitem{lin2017focal}
Lin, T.Y., Goyal, P., Girshick, R., He, K., Doll{\'a}r, P.: Focal loss for
  dense object detection. In: IEEE ICCV (2017)

\bibitem{mosinska2018beyond}
Mosinska, A., Marquez-Neila, P., Kozi{\'n}ski, M., Fua, P.: Beyond the
  pixel-wise loss for topology-aware delineation. In: IEEE CVPR (2018)

\bibitem{mou2021cs2}
Mou, L., Zhao, Y., Fu, H., Liu, Y., Cheng, J., Zheng, Y., Su, P., Yang, J.,
  Chen, L., Frangi, A.F., et~al.: Cs2-net: Deep learning segmentation of
  curvilinear structures in medical imaging. Medical image analysis
  \textbf{67},  101874 (2021)

\bibitem{ronneberger2015u}
Ronneberger, O., Fischer, P., Brox, T.: U-net: Convolutional networks for
  biomedical image segmentation. In: International Conference on Medical image
  computing and computer-assisted intervention. pp. 234--241. Springer (2015)

\bibitem{sasaki2017joint}
Sasaki, K., Iizuka, S., Simo-Serra, E., Ishikawa, H.: Joint gap detection and
  inpainting of line drawings. In: IEEE CVPR (2017)

\bibitem{schapire1999brief}
Schapire, R.E.: A brief introduction to boosting. In: IJCAI. vol.~99, pp.
  1401--1406. Citeseer (1999)

\bibitem{schroff2015facenet}
Schroff, F., Kalenichenko, D., Philbin, J.: Facenet: A unified embedding for
  face recognition and clustering. In: Proceedings of the IEEE conference on
  computer vision and pattern recognition. pp. 815--823 (2015)

\bibitem{shit2021cldice}
Shit, S., Paetzold, J.C., Sekuboyina, A., Ezhov, I., Unger, A., Zhylka, A.,
  Pluim, J.P., Bauer, U., Menze, B.H.: cl{D}ice-a novel topology-preserving
  loss function for tubular structure segmentation. In: IEEE CVPR (2021)

\bibitem{drive}
Staal, J., Abr{\`a}moff, M.D., Niemeijer, M., Viergever, M.A., Van~Ginneken,
  B.: Ridge-based vessel segmentation in color images of the retina. IEEE
  Transactions on Medical Imaging  \textbf{23}(4),  501--509 (2004)

\bibitem{wu2017fuiqa}
Wu, L., Cheng, J.Z., Li, S., Lei, B., Wang, T., Ni, D.: Fuiqa: fetal ultrasound
  image quality assessment with deep convolutional networks. IEEE transactions
  on cybernetics  \textbf{47}(5),  1336--1349 (2017)

\end{thebibliography}

\end{document}